# 3-D IR imaging with uncooled GaN photodiodes using nondegenerate two-photon absorption

Himansu S. Pattanaik,[1] Matthew Reichert,[1,3] David J. Hagan,[1,2,*] and Eric W. Van Stryland[1,2]

[1]*CREOL, The College of Optics and Photonics, University of Central Florida, Orlando, Florida 32816, USA*
[2]*Department of Physics, University of Central Florida, Orlando, Florida 32816, USA*
[3]*Current address: Department of Electrical Engineering, Princeton University, Princeton, New Jersey 08455, USA*
[*]hagan@creol.ucf.edu

**Abstract:**

We utilize the recently demonstrated orders of magnitude enhancement of extremely nondegenerate two-photon absorption in direct-gap semiconductor photodiodes to perform scanned imaging of 3D structures using IR femtosecond illumination pulses (1.6 μm and 4.93 μm) gated on the GaN detector by sub-gap, femtosecond pulses. While transverse resolution is limited by the usual imaging criteria, the longitudinal or depth resolution can be less than a wavelength, dependent on the pulsewidths in this nonlinear interaction within the detector element. The imaging system can accommodate a wide range of wavelengths in the mid-IR and near-IR without the need to modify the detection and imaging systems.

## 1. Introduction

There are numerous emerging applications of infrared (IR), and particularly mid-IR, wavelength sources since this is the molecular fingerprint spectral region exhibiting well-defined absorption bands attributed to specific absorbing molecules [1–3]. This spectral region for sensing is important for human breath analysis, tissue spectroscopy (collagens, lipids, proteins, glycogens, etc.), and for monitoring of e.g., glucose, water and air pollutants, pharmaceuticals, toxic agents, etc [1–5]. There are also important applications in the area of spectral imaging and detection of mid-IR wavelengths: optical coherence tomography has been shown in the mid-IR region for noninvasive monitoring of the structure and biochemical content of engineered tissue during the growth cycle [6], 3-D multispectral stand-off imaging is shown for target discrimination [4], mid-IR imaging is used as a relevant technique complementing near-IR imaging of biological tissues [3], etc. [2,5,7]. There is also potential interest in characterization and detection of microstructures manufactured in industrial ceramics, e.g. alumina and zirconia, for application in microfluidic devices for microreactors, fuel cells, and medical devices, etc. [8]. Su et al. [8] showed an optimal wavelength within the 1.3 μm to 6 μm range for inspection of micromanufacturing of alumina and zirconia based devices. There are two basic modes of carrying out mid-IR spectroscopy and imaging, i) transmission mode and ii) reflection mode. The reflection mode has potential for clinical diagnostic applications in imaging as it allows in-vivo operation on biological samples. There has been considerable work in near-IR and mid-IR diffuse reflectance imaging [3]. Mid-IR radiation reflected from biological tissues is scattered much less than visible light, owing to the wavelength dependence of Rayleigh-like scattering ( $\propto \lambda^{-4}$ ) of biological tissues in the mid-IR region [3]. For the near-IR region the scattering is more Mie-like, resulting in greater scattering, and hence penetrating less deeply into the biological tissues [3]. Liquid nitrogen cooled HgCdTe (MCT) detectors based on linear absorption of the mid-IR light are often used [3–7]. Apart from linear detection techniques different nonlinear methods have been tried. Frequency upconversion using the $\chi^{(2)}$ nonlinearity to up-convert IR photons to higher energy photons via sum-frequency generation to be detected by high quantum efficiency detectors is a promising technique. Recently, Dam et al. [1] showed a highly sensitive field deployable upconversion device which can acquire mid-IR images between 2.85 and ~ 5 μm. The system can acquire images of thermal light sources at room temperature and is the first mid-IR single photon imaging device applied to high-resolution spectral imaging.

Two-photon absorption (2PA), which is a $\chi^{(3)}$ process, generates a photoresponse that varies quadratically with the incident irradiance for degenerate photon energies. However, 2PA can occur through annihilation of two-photons of the same photon energy (degenerate (D)) or different photon energies (nondegenerate (ND)) to create an electron-hole pair. Reports available using D-2PA have used a photomultiplier tube (PMT) or avalanche photo diode (APD) [9,10] due to the relatively small signals. For the ND case, the photogenerated carrier density is linearly proportional to the irradiance at each frequency. We have demonstrated that using extremely nondegenerate (END) photons (energy ratio ~10), there is an orders of magnitude increase in 2PA compared to the degenerate case in direct-gap semiconductors [11–13]. Based on this enhancement, we have shown sensitive detection of mid-IR (5.6 μm) femtosecond pulses in an uncooled p-i-n GaN photodiode [14], which surpassed the response of a HgCdTe detector when using femtosecond IR in combination with 390 nm subgap femtosecond gating pulses. Details of this process are given in Ref. [13,14] and a summary is given in Sec. 2 below. The sensitivity of the method is attributed to the fact that the IR signal pulse to be detected is gated with a pulse of photon energy just below the bandgap for ND-2PA. By flooding the detector with near-gap photons, the IR photon always has a near-gap photon to pair with to create an electron and hole.

This process can be considered as the $\chi^{(3)}$ equivalent of the $\chi^{(2)}$ up-conversion technique, which has been shown to be highly efficient with conversion efficiencies, now approaching unity. However, the $\chi^{(2)}$ up-conversion method requires a separate detector and fabrication of periodically poled upconversion crystals for phase matching [15–22]. IR detection using END-2PA is simpler since the detector element itself is the active material for 2PA. But this method is new and requires considerable research and development for practical devices. Commercial detectors are far from optimized for 2PA detection.

In this paper we extend our work of mid-IR femtosecond pulse detection to show scanning 3-D mid-IR imaging in an uncooled GaN photodiode using END-2PA. The primary aim of this paper is to demonstrate the feasibility of IR imaging using this technique with uncooled wide-gap photodiodes. We used commercial p-i-n diodes which are not optimized for 2PA. To show this proof of principle, we have carried out IR imaging on metallic and semiconductor objects. The subsequent sections are organized as follows: the theory section gives a background of the detection technique; the methods section describes details of the imaging system and the process of automated data acquisition; the experimental results section shows examples of 3D images and describes both the transverse and the longitudinal resolution; the discussion and conclusion section describes the usefulness of this imaging technique for different applications and conclusions of the work carried out in this paper.

## 2. Theoretical background

Semiconductors have large third-order nonlinearities and have been subject to extensive research both theoretically and experimentally to utilize these nonlinearities for a variety of applications [23]. A simple 2-parabolic band model of semiconductors does well in describing the spectrum and magnitude of 2PA in semiconductors for both degenerate and nondegenerate photons. This theory leads to universal scaling rules that agree with data obtained for both D and ND photons [13,24,25]. The 2PA coefficient, $\alpha_2$, scales as $E_g^{-3}$, where $E_g$ is the bandgap energy, leading to larger 2PA for smaller gap materials. However, because the range of 2PA is between $E_g/2 < \hbar\omega < E_g$, as the gap decreases, the applicable wavelength range for 2PA detection gets pushed farther into the IR. These scaling rules also give the frequency dependence of 2PA for both D and ND cases.

As shown in Ref. [14], the minimum 2PA occurs for equal photon energies and can be greatly enhanced by using two very different photon energies. For example, in ZnSe, 2PA is enhanced by ≈ 270× the value in the degenerate case as measured using photon energies of 2.563 eV and 0.214 eV, i.e. a photon energy ratio of 12 [13]. This nondegenerate enhancement is due to intermediate state resonances. The END-2PA process for a direct-gap semiconductor is shown in Fig. 1, where simultaneous absorption of two photons of different energies creates an electron and hole pair. For END-2PA, when the large photon energy becomes near resonant to the bandgap (interband transition) the smaller photon energy can become nearly resonant to the intraband transition of zero frequency resonance (Fig. 1 (b)).

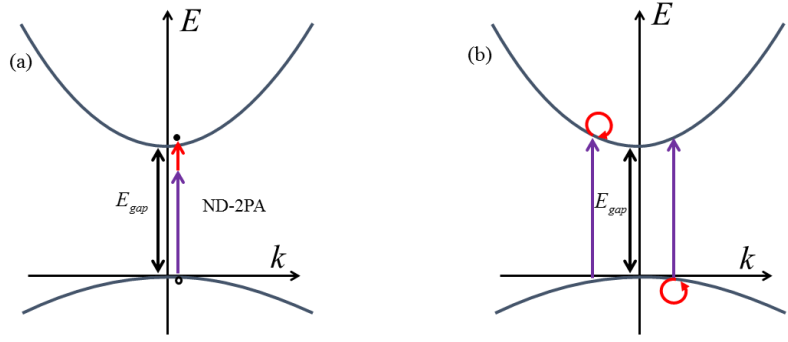

Fig. 1. (a) Schematic representation of the ND-2PA process for photons having different energies. (b) Possible transition paths in ND-2PA for two-band structure consisting of interband and intraband transitions.

The relevant equation describing the carrier generation rate from Ref. [12–14] is:

$$\frac{dN}{dt} = \frac{dN}{dt}\bigg|_{ND} + \frac{dN}{dt}\bigg|_{D} = 2\alpha_2^{ND}(\omega_s;\omega_g)\frac{I_g I_s}{\hbar\omega_s} + \alpha_2^{D}(\omega_g;\omega_g)\frac{I_g^2}{\hbar\omega_g}, \quad (1)$$

where

$$\alpha_2^{ND}(\omega_s;\omega_g) = K\frac{\sqrt{E_p}}{n_s n_g E_g^3} F_2^{ND}\left(\frac{\hbar\omega_s}{E_g};\frac{\hbar\omega_g}{E_g}\right) \quad (2)$$

and

$$\alpha_2^{D}(\omega_g;\omega_g) = K\frac{\sqrt{E_p}}{n_g^2 E_g^3} F_2^{ND}\left(\frac{\hbar\omega_g}{E_g},\frac{\hbar\omega_g}{E_g}\right) \quad (3)$$

with $F_2^{ND}(x_1;x_2) = \frac{(x_1+x_2-1)^{3/2}}{2^7 x_1 x_2^2}\left(\frac{1}{x_1}+\frac{1}{x_2}\right)^2$. Here $\alpha_2^{ND}(\omega_s;\omega_g)$ is the ND-2PA coefficient for optical frequencies $\omega_s$ and $\omega_g$, $\alpha_2^{D}(\omega_g;\omega_g)$ is the D-2PA coefficient for optical frequency $\omega_g$, $E_p$ is the Kane energy [11,12], and $K$ is a material-independent, experimentally determined constant $K = 3100$ $cm\ GW^{-1}eV^{5/2}$, which is within a factor of 26% of the two-parabolic band model prediction for D-2PA in semiconductors [25]. For IR detection, $I_g(\omega_g)$ is the irradiance of the gate beam which is at the shorter wavelength, and $I_s(\omega_s)$ is the irradiance of the signal beam which is at the longer IR wavelength [13].

The large enhancement of 2PA may be used for a variety of applications [13,14]. It is possible to monitor the photogenerated charge carriers from an intense pulse, which could be either the high energy photons or the low energy photons. In one of the applications based on this enhancement we reported pulsed mid-IR detection at 5.6 μm in uncooled GaN p-i-n photodiodes [14] where the intense gate pulse switches the detector 'on' by sensitizing the photodetector to the signal pulse via ND-2PA.

In the case of the higher energy photons being used as the gate pulses, the equation for carrier generation consists of both D-2PA due to gate photons and the ND-2PA term due to the interaction of gate and signal photons. In the END-2PA based IR detection, the photocarriers created through D-2PA present a constant background which effectively acts as an additional source of dark current. This background due to D-2PA can be discriminated against by modulating the signal pulse and using synchronous detection techniques [14]. Thus the signal detected is linear in the irradiance of each input wave. The noise present in the D-2PA signal affects the signal-to-noise (SNR) in the END-2PA based IR detection.

## 3. Experimental Methods

Imaging of backscattered IR (both near-IR and mid-IR) was performed using a commercial GaN p-i-n photodetector of bandgap energy ≈ 3.42 eV, which corresponds to a wavelength of 365 nm. The gate and signal pulses were produced using various nonlinear frequency conversion devices pumped by a regeneratively amplified, 1 kHz repetition rate, Ti:Sapphire laser system (ClarkMXR, CPA 2010) producing 780 nm femtosecond pulses (≈ 1 mJ and ≈ 150 fs (FWHM)). For example, an optical parametric amplifier (TOPAS-C, Light Conversion Ltd.) was used to obtain the signal (1100 nm – 1550 nm) and idler (1550 nm – 2630 nm) pulses. Mid-IR pulses were obtained by difference frequency generation (DFG) of the signal and idler pulses in a AgGaS crystal. In all the experiments described in this paper, the gate pulse has a wavelength of 390 nm, just below the GaN gap, which was obtained by frequency doubling a portion of the 780 nm Ti:Sapphire output using a BBO crystal. The experiments were carried out as shown in Fig. 2, where the gate and signal interact noncollinearly at an angle ~15º on the photodetector. The output voltage from the detector is due to both the ND-2PA of the gate and signal and the D-2PA of the gate. The voltage on the photodetector is only weakly dependent on the angle at which the gate pulse and the signal pulse overlap at the photodetector unlike in the detection through the phase-matched $\chi^{(2)}$ upconversion technique [1]. To detect only the IR we modulate the 1 kHz repetition rate IR pulse at 285 Hz and use synchronous detection to record the ND-2PA signal.

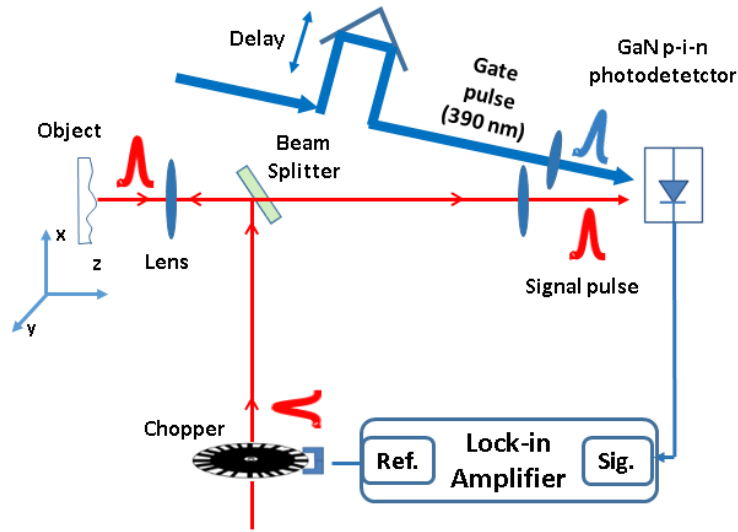

Fig. 2. Experimental configuration for single pixel scanning 3-D IR imaging.

Referring to Fig. 2, the IR pulse is focused onto the object after passing through a chopper, and the collected scattered light from the object is focused onto the GaN photodetector where it overlaps with the temporally delayed gate pulse. The GaN p-i-n photodetector used for the experiments has an active area of 0.5 mm$^2$ and an intrinsic layer of thickness 0.5 μm. To obtain the image, the object is raster scanned in the *x* and *y* directions (Fig. 2) and for each (*x*,

*y*) coordinate the ND-2PA signal from the GaN detector is measured as a function of the relative time delay between the IR pulse and the gate pulses yielding their cross-correlation. The recorded signal contains two imaging modalities which provide two different types of information about the object: 1) A 3-D image is obtained by noting the position of the peak of the cross-correlation signal for each (*x, y*) coordinate, which provides depth information about the object 2) A 2-D image is obtained by recording the photodetector output magnitude at the peak of the cross-correlation curve for each (*x, y*) coordinate, which contains information on the surface scatter.

## 4. Experimental Results

To demonstrate the technique we raster scanned several objects The signal pulse was first chosen to be in the near-IR region at 1600 nm, which has a pulse width of 111 fs, as measured by autocorrelation. We chose a reflective object as shown in Fig. 3 to have a large SNR at the photodetector. The object is a coin as shown in Fig. 3(a). To obtain the image, the object is raster scanned with a grid spacing of 250 μm × 500 μm using the 1600 nm signal pulse.

For each (*x, y*) coordinate the cross-correlation signals were recorded. The raster scanning translation stages of the object and the delay stage are automated and controlled using the same Labview software that records the detector signal to obtain the image. Fig. 3(b) shows the cross correlation curves for two different positions on the object. The *x,y* resolution is limited by the rather course imaging grid that was used.

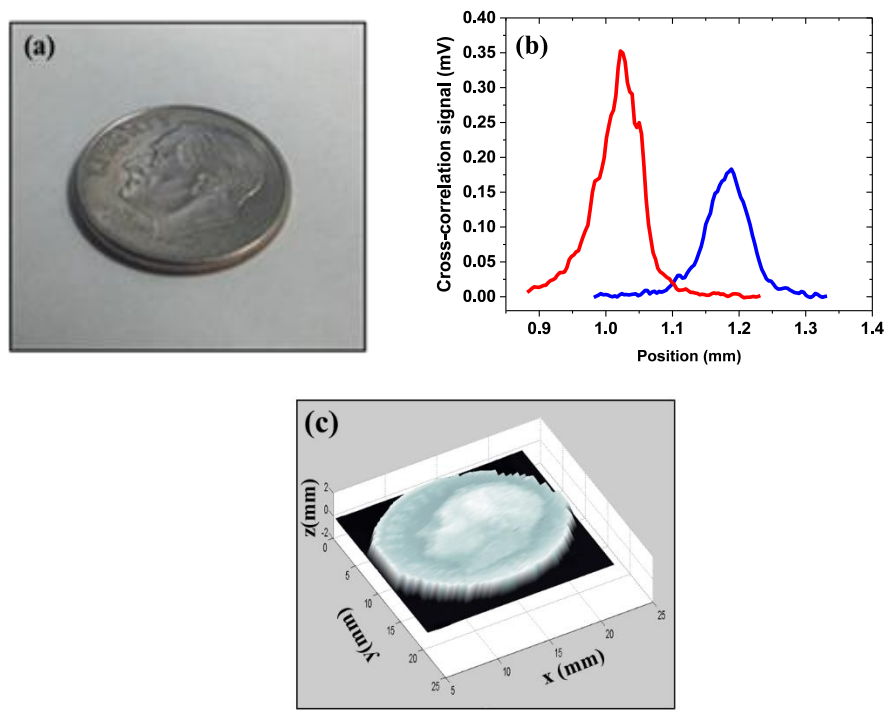

Fig. 3. (a) Photograph of the dime. (b) Cross-correlation curves for two different points on the dime. (c) 3-D image of the dime taken at 1600 nm.

A similar experimental configuration is used to carry out mid-IR imaging with a 4.93 μm signal pulse. In this setup the near IR beam is replaced by the mid-IR beam. The purpose of choosing 4.93 μm for mid-IR imaging is the low atmospheric transmission loss and the availability of a PbSe photodetector (ThorLabs, PDA20H), which we use to characterize the mid-IR beam. The 3-D image obtained in this work with the ND-2PA technique is a surface 3-D image. For the mid-IR imaging we chose an object (Fig. 4(a)) which is a steel wire-die that has numbers embossed on it ("80"). Figs. 4(b) and 4(c) show the 3-D image and 2-D image respectively, obtained from the cross-correlation curve as described before. Here the object is raster scanned with a finer grid spacing (80 μm × 80 μm). For comparison we replaced the GaN photodiode with a PbSe detector and recorded the image with 4.93 μm in staring mode shown in Fig. 4 (d).

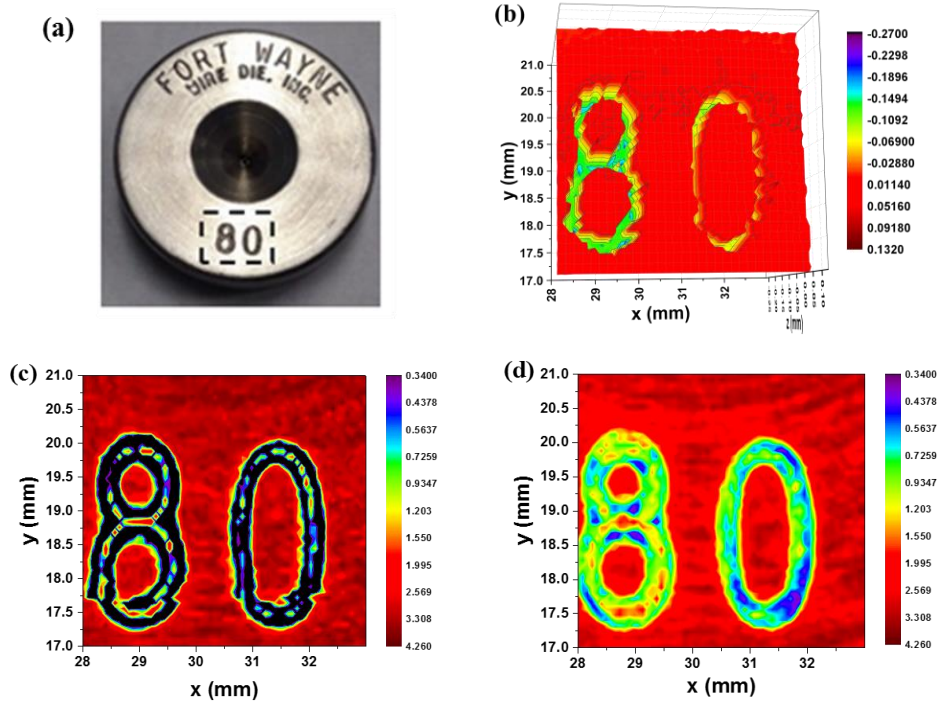

Fig. 4. (a) Photograph of the object '80'. (b) 3-D reflectance image of the object, (c) 2-D reflectance image of the object. (d) 2-D reflectance image using a PbSe detector in staring mode. The kinks in images (b) and (c) at y = 17.75 mm is due to the malfunction of the raster scanning stage.

*4.1 Resolution*

The transverse resolution in the obtained images is determined by the digitized raster scanning resolution, but the ultimate spatial resolution is determined by the minimum spot size obtained through the imaging optics for the signal pulse. The longitudinal (depth) resolution is determined by the accuracy of measuring the peak delay positions for the cross correlations. The error in this determination is limited by noise in the cross-correlation curve.

Considering the cross-correlation curves to be Gaussian the output cross-correlation signals are represented as $I = I_0 \exp(-t/\tau_{FWHM})^2$ with zero noise, and $I = I_0 (1+\sigma)\exp(-t/\tau_{FWHM})^2$ with noise σ. The error in determination of zero delay is obtained as

$$\Delta z = \pm \frac{1}{4\sqrt{\ln 2}} \sqrt{\ln(1+\sigma)} \tau_{FWHM} c, \qquad (4)$$

where $\tau_{FWHM}$ is the full width half maximum (FWHM) of the cross-correlation curve, $c$ is the velocity of light. In the experiments the noise on the cross-correlation curves are ≈ 2% and the FWHM of the cross-correlation curve ≈ 340 fs. From Eq. (4), for noise σ = 2% on the cross-correlation signal gives $\Delta z \approx \pm 4$ µm. The above assumes that the depth is well-defined within the focal spot. In that case, the estimation of depth resolution obtained using Eq. (4) is a conservative value as better depth resolution can be obtained by curve fitting each of the cross-correlation curves. On the other hand, for a reflective surface, knowing the individual pulsewidths, the width of the measured cross correlation gives a measure of the surface roughness. In the experiments performed here the cross correlation widths from the samples were the same as the width when reflected off a mirror.

*4.2 Mid-IR imaging of buried structures*

Imaging of laser written volumetric structures or any other buried structures could be obtained using this technique provided the substrate material is transparent to the wavelength of the signal pulse. To demonstrate the image acquisition of buried structures we chose the object to be a GaAs semiconductor structure (Fig. 5(a)) of depth ≈ 21 µm as measured by a profilometer (alpha-step 200, Tencor Instruments) The image is obtained by raster scanning the structure with light incident from the substrate side ( Fig. 5(b))

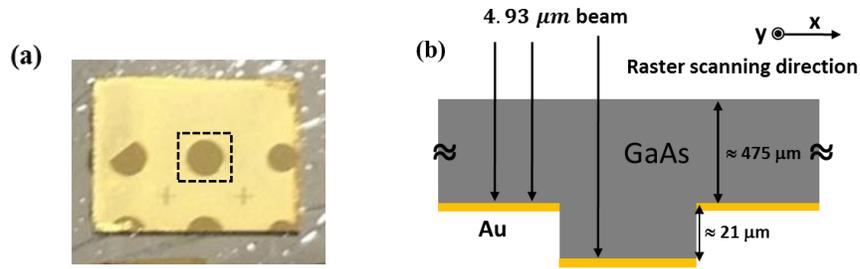

Fig. 5. (a) Photograph of the GaAs semiconductor structure. The squared region shows the raster scanned area (b) Sketch describing the raster scanning carried out from the substrate side.

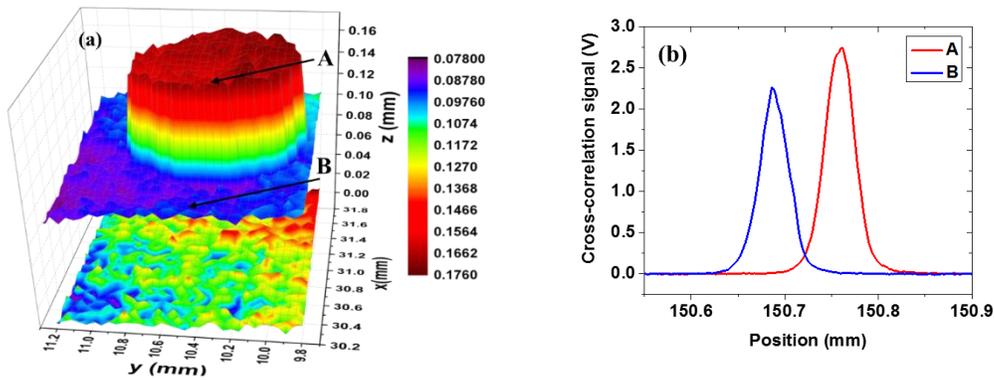

Fig. 6. (a) 3-D image of the portion of the structure scanned for the 3-D imaging. (b) Cross-correlation signal corresponding to the position 'A' and 'B'.

The 3-D image obtained (Fig. 6) shows the optical path length within the structure is 69 ± 6 µm. The height of the structure then obtained by dividing the optical path length by the refractive index of the GaAs at 4.93 µm, which gives the height of the structure 21 ± 2 µm. The depth resolution is less than the vacuum wavelength used. This

'superresolution' in depth is due to the ultrafast temporal gating detection method used and is limited by the pulsewidth and SNR in the cross-correlation data. This enables us to obtain post-fabrication images of buried structures such as the semiconductor structures or any laser written volumetric structures as described elsewhere [26].

## 7. Discussion and Conclusion

The method described here is an experimental investigation of how the END-2PA in wide-gap semiconductor photodetectors can be utilized for imaging in the mid-IR as well as in the near IR spectral region. In our previous work [14] we showed the sensitivity and signal-to-noise ratio (SNR) of the gated detection of mid-IR femtosecond pulses is comparable to that of commercially available HgCdTe (MCT) detectors, while using unoptimized commercial photodiodes. The GaN p-i-n photodetector used for the current experiments are still not optimized, and has an intrinsic layer thickness of only 0.5 µm. For END-2PA detection, thicker elements are desirable for longer interaction of the gate and signal pulses. The design of such thick p-i-n detector elements is the subject of ongoing research to make more efficient detectors for END-2PA. One possible solution is to make the gate and signal beam interact longer using a waveguide p-i-n photodetector.

In addition to the samples shown, the technique described here could be useful for imaging of biological samples. Previous work [2–6] in the area of mid-IR imaging of biological specimens used CW or nanosecond pulses in the mid-IR region.

Another useful application, as described in the experimental setup section, is the imaging of buried structures in bulk IR transmissive materials such as semiconductors and certain ceramics. Nondestructive 3-D imaging of laser-written volumetric structures could be performed provided the substrate is transparent to the incident IR beam. In Ref. [26] the authors showed nondestructive 3-D imaging using OCT of the laser-written structures. The technique facilitates real time monitoring of the 3-D structures created in fused silica and hence provides quantitative assessment of the features for better control of the writing process. Our technique could be useful for similar applications. As an example, we have shown here post fabrication imaging of a semiconductor device structure.

We demonstrated the feasibility of mid-IR imaging in an uncooled wide-gap photodetector using END-2PA. The 3-D and 2-D images obtained give a proof-of-principle of the mid-IR imaging capability of END-2PA detection at room temperature. The potential for sub-wavelength resolution was presented. Assuming depth variations within the spot size are negligible, the depth resolution is determined by the cross-correlation width. The method described here is nondestructive and suitable for in-vivo applications in biology. There are many avenues for significant improvement to reach the full potential of this technique. An advantage of this system is that the imaging can be carried out at a wide variety of wavelengths in the mid-IR and near-IR without any modification to the detection and imaging systems.

**Acknowledgements**

This work was supported by the National Science Foundation grants ECCS-1202471 and ECCS-1229563.

**References**

1. J. S. Dam, P. Tidemand-Lichtenberg, and C. Pedersen, "Room-temperature mid-infrared single-photon spectral imaging," Nat. Photonics **6**, 788–793 (2012).

2. A. B. Seddon, "Mid-infrared (IR) - A hot topic: The potential for using mid-IR light for non-invasive early detection of skin cancer in vivo," Phys. Status Solidi **250**, 1020–1027 (2013).

3. B. Guo, Y. Wang, C. Peng, H. Zhang, G. Luo, H. Le, C. Gmachl, D. Sivco, M. Peabody, and a Cho, "Laser-based mid-infrared reflectance imaging of biological tissues.," Opt. Express **12**, 208–19 (2004).

4. Y. Wang, Y. Wang, and H. Q. Le, "Multi-spectral mid-infrared laser stand-off imaging," Opt. Express **13**, 6572–6586 (2005).

5. M. J. Walsh, R. K. Reddy, R. Bhargava, and A. Member, "Label-Free Biomedical Imaging With Mid-IR Spectroscopy," IEEE J. Sel. Top. Quantum Electron. **18**, 1502–1513 (2012).

6. C. S. Colley, J. C. Hebden, D. T. Delpy, A. D. Cambrey, R. A. Brown, E. A. Zibik, W. H. Ng, L. R. Wilson, and J. W. Cockburn, "Mid-infrared optical coherence tomography.," Rev. Sci. Instrum. **78**, 123108 (2007).

7. S. Liakat, A. P. M. Michel, K. A. Bors, and C. F. Gmachl, "Mid-infrared ($\lambda$ = 8.4–9.9 µm) light scattering from porcine tissue," Appl. Phys. Lett. **101**, 093705 (2012).

8. R. Su, M. Kirillin, E. W. Chang, E. Sergeeva, S. H. Yun, and L. Mattsson, "Perspectives of mid-infrared optical coherence tomography for inspection and micrometrology of industrial ceramics.," Opt. Express **22**, 15804–19 (2014).

9. J. M. Roth, T. E. Murphy, and C. Xu, "Ultrasensitive and high-dynamic-range two-photon absorption in a GaAs photomultiplier tube.," Opt. Lett. **27**, 2076–8 (2002).

10. F. Boitier, J.-B. Dherbecourt, A. Godard, and E. Rosencher, "Infrared quantum counting by nondegenerate two photon conductivity in GaAs," Appl. Phys. Lett. **94**, 081112 (2009).

11. D. C. Hutchings and E. W. Van Stryland, "Nondegenerate two-photon absorption in zinc blende semiconductors," J. Opt. Soc. Am. B **9**, 2065–2074 (1992).

12. M. Sheik-bahae, D. C. Hutchings, D. J. Hagan, E. W. Van Stryland, and S. Member, "Dispersion of Bound Electronic Nonlinear Refraction in Solids," IEEE J. Quantum Electron. **27**, 1296–1309 (1991).

13. C. M. Cirloganu, L. A. Padilha, D. A. Fishman, S. Webster, D. J. Hagan, and E. W. Van Stryland, "Extremely nondegenerate two-photon absorption in direct-gap semiconductors [Invited].," Opt. Express **19**, 22951–60 (2011).

14. D. A. Fishman, C. M. Cirloganu, S. Webster, L. A. Padilha, M. Monroe, D. J. Hagan, and E. W. Van Stryland, "Sensitive mid-infrared detection in wide-bandgap semiconductors using extreme non-degenerate two-photon absorption," Nat. Photonics **5**, 561–565 (2011).

15. R. H. Hadfield, "Single-photon detectors for optical quantum information applications," Nat. Photonics **3**, 696–705 (2009).

16. R. T. Thew, H. Zbinden, and N. Gisin, "Tunable upconversion photon detector," Appl. Phys. Lett. **93**, 071104 (2008).

17. J. Pelc, C. R. Phillips, C. Langrock, Q. Zhang, L. Ma, O. Slattery, X. Tang, and M. M. Fejer, "Single-Photon Detection at 1550 nm via Upconversion Using a Tunable Long-Wavelength Pump Source," CLEO2011 - Laser Appl. to Photonic Appl. CMC4 (2011).

18. G.-L. Shentu, J. S. Pelc, X.-D. Wang, Q.-C. Sun, M.-Y. Zheng, M. M. Fejer, Q. Zhang, and J.-W. Pan, "Ultralow noise up-conversion detector and spectrometer for the telecom band.," Opt. Express **21**, 13986–91 (2013).

19. J. S. Pelc, L. Ma, C. R. Phillips, Q. Zhang, C. Langrock, O. Slattery, X. Tang, and M. M. Fejer, "single-photon detector at 1550 nm : performance and noise analysis," Opt. Express **19**, 1725–1727 (2011).

20. P. S. Kuo, O. Slattery, Y.-S. Kim, J. S. Pelc, M. M. Fejer, and X. Tang, "Spectral response of an upconversion detector and spectrometer.," Opt. Express **21**, 22523–31 (2013).

21. M. A. Albota and F. N. C. Wong, "Efficient single-photon counting at 1 . 55 micron by means of frequency upconversion," Opt. lett **29**, 1449–1451 (2004).

22. X. Gu, K. Huang, Y. Li, H. Pan, E. Wu, and H. Zeng, "Temporal and spectral control of single-photon frequency upconversion for pulsed radiation," Appl. Phys. Lett. **96**, 131111 (2010).

23. D. N. Christodoulides, I. C. Khoo, G. J. Salamo, G. I. Stegeman, and E. W. Van Stryland, "Nonlinear refraction and absorption: mechanisms and magnitudes," Adv. Opt. Photonics **2**, 60 (2010).

24. B. S. Wherrett, "Scaling rules for multiphoton interband absorption in semiconductors," J. Opt. Soc. Am. B **1**, 67 (1984).

25. E. W. Van Stryland, M. A. Woodall, H. Vanherzeele, and M. J. Soileau, "Energy band-gap dependence of two-photon absorption," Opt. Lett. **10**, 490 (1985).

26. J. Choi, K.-S. Lee, J. P. Rolland, T. Anderson, and M. C. Richardson, "Nondestructive 3-D imaging of femtosecond laser written volumetric structures using optical coherence microscopy," Appl. Phys. A **104**, 289–294 (2010).